\documentclass[sigconf]{acmart}

\AtBeginDocument{%
  \providecommand\BibTeX{{%
    \normalfont B\kern-0.5em{\scshape i\kern-0.25em b}\kern-0.8em\TeX}}}

\setcopyright{none}
\copyrightyear{2021}
\acmYear{2021}
\acmDOI{10.1145/3489849.3489871}

\acmConference[VRST'21]{VRST '21: ACM Symposium on Virtual Reality Software and Technology }{December 08--10, 2021}{ Osaka, Japan}
\acmBooktitle{VRST '21: ACM Symposium on Virtual Reality Software and Technology,
  December 08--10, 2021,  Osaka, Japan}
\acmPrice{15.00}
\acmISBN{978-1-4503-XXXX-X/18/06}

\graphicspath{{Images/}}

\usepackage{lscape}

\usepackage{acronym}

\acrodef{ve}[VE]{Virtual Environment}
\acrodef{vr}[VR]{Virtual Reality}
\acrodef{ris}[RIS]{Real-time Interactive System}
\acrodef{dve}[DVE]{Distributed Virtual Environment}
\acrodef{cve}[CVE]{Collaborative Virtual Environment}
\acrodef{mvr}[MVR]{Mobile Virtual Reality}
\acrodef{rtt}[RTT]{Round-Trip Time}
\acrodef{qoe}[QoE]{Quality of Experience}
\acrodef{snr}[SNR]{Signal to Noise Ratio}
\acrodef{psd}[PSD]{Power Spectral Density}
\acrodef{rmse}[RMSE]{Root Mean Squared Error}
\acrodef{dof}[DOF]{Degrees of Freedom}
\acrodef{cbn}[CBN]{Consensus Based Networking}
\acrodef{pbd}[PBD]{Position Based Dynamics}
\acrodef{pcam}[PCAM]{Predictive Contract Agreement Mechanism}
\acrodef{qos}[QoS]{Quality of Service}
\acrodef{ptp}[PTP]{Precision Time Protocol}
\acrodef{evd}[EVD]{Eigen Vector Decomposition}
\acrodef{pdes}[PDES]{Parallel Discrete Event Simulation}
\acrodef{ekf}[EKF]{Extended Kalman Filter}
\acrodef{pid}[PID]{Proportional-Integral-Derivative}
\acrodef{guid}[GUID]{Globally Unique Identifier}

\usepackage{subcaption}

\begin{document}

\title{Ubiq: A System to Build Flexible Social Virtual Reality Experiences}


\author{Sebastian Friston}
\email{sebastian.friston@ucl.ac.uk}
\author{Ben Congdon}
\email{ben.congdon.11@ucl.ac.uk}
\author{David Swapp}
\email{d.swapp@ucl.ac.uk}
\affiliation{%
  \institution{University College London}
  \country{United Kingdom}
}

\author{Lisa Izzouzi}
\email{l.izzouzi@ucl.ac.uk}
\author{Klara Brandst\"{a}tter}
\email{k.brandstatter@ucl.ac.uk}
\author{Daniel Archer}
\email{daniel.archer.18@ucl.ac.uk}
\affiliation{%
  \institution{University College London}
  \country{United Kingdom}
}

\author{Otto Olkkonen}
\email{otto.olkkonen.20@ucl.ac.uk}
\author{Felix J. Thiel}
\email{felix.thiel.18@ucl.ac.uk}
\author{Anthony Steed}
\email{a.steed@ucl.ac.uk}
\affiliation{%
  \institution{University College London}
  \country{United Kingdom}
}

%


\renewcommand{\shortauthors}{Friston, Congdon, et al.}

\begin{abstract}
While they have long been a subject of academic study, social virtual reality (SVR) systems are now attracting increasingly large audiences on current consumer virtual reality systems. The design space of SVR systems is very large, and relatively little is known about how these systems should be constructed in order to be usable and efficient. In this paper we present Ubiq, a toolkit that focuses on facilitating the construction of SVR systems. We argue for the design strategy of Ubiq and its scope. Ubiq is built on the Unity platform. It provides core functionality of many SVR systems such as connection management, voice, avatars, etc. However, its design remains easy to extend. We demonstrate examples built on Ubiq and how it has been successfully used in classroom teaching. Ubiq is open source (Apache License) and thus enables several use cases that commercial systems cannot.
\end{abstract}




\begin{teaserfigure}
\centering
  \includegraphics[width=0.73\textwidth]{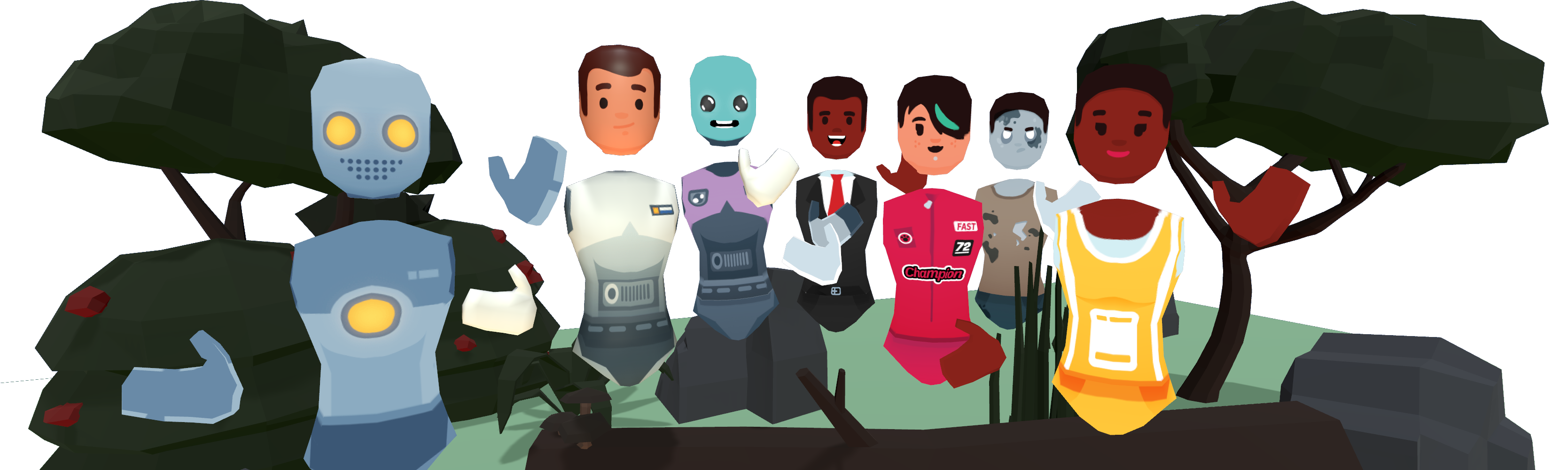}
  \Description{Seven stylised avatars waving at the camera in the Hello World environment. The environment consists of similarly stylised logs and trees above a green ground plane.}
  \caption{Ubiq's social sample application showing cartoony floating avatars}
  \Description{}
  \label{fig:teaser}
\end{teaserfigure}

\maketitle

\section{Introduction}

In the past five years, social virtual reality systems (SVRs) have emerged as one of the most promising applications of consumer VR. SVRs are characterised by a relatively unstructured experience of meeting and socialising with other users. While many SVRs provide game-like environments to explore, the focus is not necessarily on play. At the time of writing, one leading system, Rec Room, announced it had more than a million monthly active VR users~\citep{hayden_rec_2021}. 

SVRs, or collaborative virtual environments (CVEs) have a history back to at least the 1990s \citep{blanchard_reality_1990, Stone1993SocialEnvironments, Mantovani1995VirtualSelves, damer_avatars_1997}. Common to almost all CVEs is some sort of representation of the users to each other as avatars. These avatars convey referencing to shared objects \citep{Hindmarsh2000Object-FocusedEnvironments} and non-verbal communication \citep{fabri_emotional_1999} in ways that are difficult in video-based media. 
However, different SVRs \& CVEs vary significantly in the types of avatars they present, the forms of their 3D user interfaces, and the functionality afforded by the environment. Recent surveys inventory the diverse design choices made \citep{jonas_towards_2019,Kolesnichenko2019UnderstandingEcology} or compare the support for specific collaborative tasks \citep{liu_social_2021}. 

The systems and infrastructure behind SVRs are also quite diverse. There have been many research systems and proposed protocols (e.g. see review in \citep{steed_networked_2009}, and Section \ref{sec:networked}), but little standardisation. Early systems were often monolithic; the networking system was bound tightly into a framework that supported the full range of capabilities needed, such as rendering, tracking and interaction. Today, developers typically access those capabilities through one of a small number of game engines such as Unity or Unreal (see Section \ref{sec:toolkits}). Indeed, access to modern consumer VR hardware is facilitated through packages on one or more of these engines. Complementary to this are toolkits that support networking, but these are often low-level and rely on external services (see Section \ref{sec:networked}).

Any SVR application needs to support a range of different functions. Thus building novel applications and demonstrations from scratch involves significant effort. This makes it hard for small teams such as student or researcher teams to build experimental systems. In this paper, we introduce Ubiq, an open source platform for prototyping SVRs. Ubiq provides a framework for developing SVRs within Unity, along with examples of common application styles and code for back-end services so that the whole system can be run without third-party services. This is especially important for applications with data protection constraints, due to ethical considerations or because of the use of proprietary data.

From a user perspective, Ubiq supports common features such as avatar selection, voice communication, shared synchronisation of objects and other features typical of such systems. From a developer perspective, Ubiq provides a simple framework and examples for building practical SVR applications. It is simple to add new types of shared objects through built-in support of some abstract concepts such as servers, rooms and messages. However, these are all well documented so that more experienced developers can extend them. Ubiq also provides a server that is easy to customise and extend.

After reviewing related work, we describe our design requirements (Section \ref{sec:designreq}) and key architectural decisions (Section \ref{sec:architecture}). In Section \ref{sec:examples}, we present a range of examples, from the default, no-configuration-necessary SVR example to some technical feature examples. In Section \ref{sec:evaluation} we present a variety of evaluations, including a feature comparison to other SVRs, technical analyses of performance, and further examples from using Ubiq in the classroom. We conclude with some future work including a preview of the feature roadmap and provision of training materials.

\section{Related Work}

\subsection{Social VR Systems}
\label{sec:svr}

One of the most compelling uses of immersive VR is as a collaborative medium. Because the user is tracked, there is a natural way to represent the user as a 3D avatar. This can then be shared with other users with a networking strategy (see Section \ref{sec:networked}). Some of the earliest VR systems had collaborative, social demonstrations (e.g. Reality Built for Two system from VPL Research \citep{blanchard_reality_1990}). There were many demonstrations of the potential of this technology in the early wave of academic VR systems \citep{Stone1993SocialEnvironments, Mantovani1995VirtualSelves, damer_avatars_1997, churchill_collaborative_1998, Latoschik2006}. There is a large body of work around avatars and the social responses that they generate \citep{biocca_toward_2003, schroeder_being_2010, Kolesnichenko2019UnderstandingEcology, latoschik_effect_2017}. Some particular features of interest are how users exploit spatial positioning relative to each other and objects \citep{Hindmarsh2000Object-FocusedEnvironments} and how users use non-verbal behaviours \citep{fabri_emotional_1999, yee_unbearable_2007, pan_impact_2017}. 

Recent interest in consumer VR has led to development of a plethora of new platforms. Schulz's blog about social VR \citep{schulz_comprehensive_2020} lists over 150 platforms. Over 250 systems are listed in the XR Collaboration directory \citep{xr_ignite_interactive_2021}. These new platforms cover a wide range of activities. While many are purely focused on gaming, a range of other application has emerged including collaborative design \cite{Wendrich2016}, training \citep{Gunn2005} and teleoperation \cite{Hokayem2006} (see also the review in \citep{dalton_reality_2021}). Recent academic surveys have started to tease apart some of the distinctive aspects of these platforms \citep{jonas_towards_2019}, the types of avatar supported \citep{Kolesnichenko2019UnderstandingEcology} or specific functionality such as facial expressions \citep{tanenbaum_how_2020}. 
Liu \& Steed \cite{liu_social_2021} presented a comparative analysis of a number of popular social VR systems, which we use to evaluate Ubiq's features in Section \ref{sec:userlevel}.

While there are a lot of platforms, a body of recent work highlights many avenues for future exploration to support their emerging uses. These include support of long-term relationships in social VR \citep{Moustafa2018AReality}, the impact of avatar representations on trust formation \citep{pan_impact_2017}, requirements for harassment prevention \citep{blackwell_harassment_2019} and users perception of their own avatars \citep{Freeman2020MyReality}.

\subsection{Networked VR}
\label{sec:networked}

The networking strategies behind SVR systems are various. As a class of technologies, networked VR is slightly broader than social VR systems as individual systems might themselves be distributed (e.g. cluster rendering~\cite{Vob2002}), or the system might draw on a variety of services hosted on the Internet \citep{singhal_networked_1999, steed_networked_2009}.  In general though, networked VR systems are concerned with sharing consistent virtual worlds in real-time.

Early networked VR systems, such as DIVE~\cite{Carlsson1993}, MASSIVE~\cite{Greenhalgh1995a} and Blue-C~\cite{Naef2003} already supported common SVR features.
This included support for heterogeneous systems, avatars, action-based interaction, message passing and spatially mediated interaction.
Such systems were built with the explicit goal of collaboration. Early work focused on network architectures, resource distribution and ownership models.
For example, one approach to constructing systems is scene-synchronisation; a common strategy in networked VR as the scene graph is a commonality among diverse implementations \cite{Drolet2009,Roth2004,Buttolo1997,Naef2003,Zeleznik2000,Latoschik2006,Allard2004}.

%
%
%
Margery et al \cite{Margery1999} classified collaboration into three levels, with L3 being the highest and referring to multiple users manipulating the same degree of freedom at the same time.
Grimstead et al's review~\cite{Grimstead2005} breaks down how many systems of the time (2005) fit into the above categories of access control, architecture and synchronisation. The review showed a definite preference for client-server architectures.
Other reviews (e.g. \cite{Staadt2003}) focus on specific applications such as cluster based rendering, which have application-specific problems (e.g. culling).


\subsection{Toolkits}
\label{sec:toolkits}

Apart from the original vertically-integrated platforms, a number of toolkits have been made for building CVEs.
These began with inherently distributed scene graphs and middleware (e.g. \cite{MacIntyre1998, Chardavoine2005, Lake2010, Humphreys2001, Hesina1999}), which provided APIs that integrated seamlessly into existing programming models.
Later academic toolkits explored different models such as modular~\cite{Latoschik2006} or component~\cite{Kharitonov2013} synchronisation, scene graph abstraction~\cite{Zeleznik2000}, and event-based state distribution~\cite{latoschik_simulator_2011}.
Some recent toolkits have been created for specific applications, such as visualisation~\cite{Dupont2010} or structured collaboration~\cite{He2020}.
However, it is still common for research projects to resort to building their own vertically integrated systems (e.g. \cite{Latoschik2019, Gugenheimer2017, Zhang2018a}).

In the commercial realm, there are a number of networking SDKs, even just considering our target platform, Unity (e.g. DarkRift or Photon; see our comparative analysis in Section \ref{sec:technical}). These SDKs are typically mid-to-low level, focusing on matchmaking and message passing. Mirror is the highest-level SDK, synchronising transforms and animations out of the box.
Despite being commercial, many SDKs are free.
However, being game-oriented, they typically make strong assumptions. For example, they are usually client-server both in architecture and authority model, and high level features don't include logging.
Low level SDKs often do not include systems such as voice chat, which are complex to implement.

Finally, it is common nowadays to build VR experiments on existing game engines, and some of these engines have networking support.
For example Unreal is inherently networked and this is reflected in its gameplay logic and state API.
Unity currently has no standard networking API, though it has an experimental low-level API in recent versions (MLAPI, see Section \ref{sec:technical})
Amazon's Lumberyard provides perhaps the highest level API of all such systems, including state synchronisation but also common game services such as leader boards.

\section{Design Requirements}
\label{sec:designreq}
\subsection{Strategic}
\label{sec:strategic}

Ubiq was designed to meet three goals that are challenging to fulfil with commercial SVRs:

\paragraph{Support Teaching VR} This includes networked and social VR. To do so effectively, a framework should be easy to learn, transparent, and integrate well with familiar tools.
It should not force students to learn concepts that are not directly relevant to the taught material.

\paragraph{Support Research in VR} This includes distributed experiments~\cite{steed2021lessons}. The platform should provide basic features that `just work' so researchers can work on their experiment rather than the platform. Non-traditional features such as logging and asymmetrical capabilities are usually required. Research also implies data flow transparency for compliance with regulations such as the CCPA or GDPR.

\paragraph{Support Research into Networked VR} Networked VR itself is an active research topic that requires platforms to experiment with. This implies transparency with the ability to modify the system and the flexibility to support different architectures and configurations.

%




\subsection{Platform Analysis}


\begin{table*}[!htp]\centering
\scriptsize
\caption{High-level analysis of main features}\label{tab:analysis}
\vspace{-0.4cm}
\begin{tabular}{lrrrrrrr}\toprule
&RecRoom &Altspace.VR &Mozilla Hubs &Spatial &VRChat &Ubiq \\\midrule
Accessibility & Create rooms \& actions & Create rooms & Create rooms/modify code & Create rooms & Create rooms \& actions &Create rooms \& code \\
Licence &Commercial &Commercial &Mozilla &Commercial &Commercial &Apache \\
Self-Hosted &No &No &Yes &No &No &Yes \\
Avatars &Customise cartoony &Customise cartoony &Cartoony/Flexible &Photo-based &Flexible &Cartoony/Rocketbox \\
Voice \& Indication &Spatialised/Icon &Spatialised/Icon &Spatialised/Animated &Spatialised/Icon &Spatialised/Icon/Animation &Spatialised/Icon \\
Instrumentation &No &No &No but see \citep{williamson_proxemics_2021} &No &No but see \citep{saffo_remote_2021} &Yes \\
Multiple Architectures &CS only &CS only &CS only &CS only &CS only &Flexible \\
\bottomrule
\end{tabular}

\end{table*}

Fulfilling these requirements with commercial platforms is challenging as these often have one or more conflicting goals. For example, security considerations prevent most platforms executing arbitrary code, which would undermine building experiments. Trade-secrets are protected through keeping source closed. 

In early stages of planning, we performed an analysis of a variety of platforms to see if there was an obvious platform to adopt. A summary of some key findings can be found in Table~\ref{tab:analysis}, which is intended to convey indicative features rather than be comprehensive. All the platforms provide ways to create new spaces, but these vary widely in the types of behaviour that can be added, from simple triggers of actions (VRChat), through to custom functionality through modified clients (Mozilla Hubs). Ubiq is at a slightly different level and is targeted at experimentation, so the functionality of Unity is available to modify environments. Mozilla Hubs and Ubiq are open source, and they both allow developers to self-host servers.

A recent taxonomy \citep{jonas_towards_2019} investigates SVR applications and emphasises novel design choices. All the considered platforms provide avatars to be embodied, but support for different avatars varies, with several supporting customisation with a style, and others (e.g. VRChat and Mozilla Hubs) permitting the import of custom models. Avatar customisation can be valuable to virtual communities to facilitate user expression and freedom, though some platforms deliberately chose more neutral avatars to prevent harassment within the VE~\citep{kolesnichenko_understanding_2019}. Ubiq provides at least two different styles of avatars.

All of the platforms provide voice, but have different ways of indicating speaking. The commercial platforms do not provide instrumentation, though it is possible to extract data via other means such as screen capture \citep{saffo_remote_2021}.

Instrumentation could be added to Mozilla Hubs (e.g. \citep{williamson_proxemics_2021}). Ubiq includes extensible instrumentation. Most platforms support client-server (CS) configurations. Only Ubiq supports flexible architectures as it is posed more as a toolkit with examples. See Section \ref{sec:technical} for a technical comparison to some other toolkits. 

While there are significant advantages to commercial platforms, in that they are well-supported and have vibrant communities, strategic needs drive us towards the type of architecture and support that Ubiq offers. Mozilla Hubs is very promising and was strongly considered for our goals. However, it is a moderately complex platform and we wanted the development experience of an IDE such as Unity. Thus, Ubiq is designed as a Unity package with samples that provide basic social VR functionality that can built upon.

\section{Architecture}
\label{sec:architecture}

\subsection{Overview}

\paragraph{Framework}

Ubiq is a framework for building SVR applications and experiments.
It consists predominantly of Unity Components, which can be used to build Unity scenes with SVR functionality, and code for a server.
Users integrate Ubiq by importing the source into their Unity projects.

\paragraph{Unity} 

We opted for Unity as it is one of the most accessible platforms for VR research and teaching. It has an inbuilt XR API so we do not have to code against individual XR SDKs. Unity does not have a built in high-level networking strategy, so there is scope to use a variety of mechanisms and conventions.

\paragraph{Project Structure}

The framework has a set of core components, and a small set of dependent samples.
The samples are not just documentation support, but contain significant functionality. The integrated SVR is implemented as a sample.
%
%
The motivation is to allow building functionality that requires some assumptions to be made, without creating dependencies in the framework.
Finally, as a fully working SVR, the samples provide a working starting point for applications.

\subsection{Messaging}
\label{sec:messaging}

\paragraph{Component-Centric Programming}
Ubiq is based around the exchange of discrete messages directly between \textit{Components}. These are instances of classes (typically Unity Components) that implement a method to receive messages.
Users implement networked behaviour in these classes. This is designed to approximate the programming model of Unity and be familiar to existing users.


\paragraph{Peers and Connections}
A Ubiq network is made up of a set of Peers. 
Each Peer may contain many Components, and many connections to other Peers.
Messages are delivered directly between components regardless of the underlying connection architecture.
For example, Peers could form a peer-to-peer mesh by each creating a connection to all others, or communicate via relay in a star arrangement.

Components can implement different logical models by controlling which other Components they address. For example, by having two asymmetrical Components exchange messages to implement a client-server model. 
For typical shared object code, the API begins and ends in the Component itself. Users do not have to write any code outside of their new Component class.

The highly abstract messaging layer prevents reliance on architectural details that are often baked into the API in other frameworks.
It clearly separates the domains of student work/VR research and networking research and allows exploring different networking technology without changing existing applications.
A disadvantage is that new users can't immediately rely on the expected client-server model, and are forced to consider what model they need for their Component.
We try to ameliorate this by providing simple examples.
Ultimately, we consider it a necessary trade-off for maintaining flexibility.

Rendezvous over the public internet does require a fixed service however.
Ubiq uses a client-server architecture for its Rooms system, where each peer makes one connection to a pre-defined server, and the server forwards messages between all peers in a Room (see Section~\ref{sec:server}).

\paragraph{Addressing}

\begin{figure}[h]
\includegraphics[width=7.5cm]{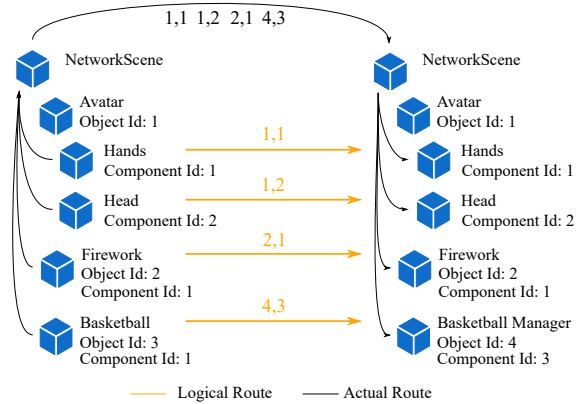}
\Description{Two scene graph diagrams side by side. The nodes of each are annotated with a distinct Id. Arrows show dataflow between nodes on the left and corresponding nodes on the right, through the root nodes of each graph.}
\caption{Diagram showing how Component instances may address each other, with the actual route taken by the messages.}
\label{fig:addressing}
\vspace{-0.5cm}
\end{figure}

Message addresses have two parts: Object Id and Component Id. Object Id is analogous to a Unity GameObject and Component Id is analogous to a Unity Component type.
Together these distinguish which individual Component instance(s) should receive a message.
Ids can be defined in different places in the scene graph. Message exchange is illustrated in Figure~\ref{fig:addressing}.

Messages are received by all Components on all Peers that match both the Object Id and Component Id. That is, the network is responsible for \textit{fanout}.
Controlling these Ids controls message delivery.

For example, symmetrical Components such as Avatars would use the same classes (Component Ids) to send and receive. 
The Object Ids for a Player's Avatar would be identical across all Peers. When the Player's Components sent updates, they would broadcast to all other Peers, without the Avatar or Player needing to know anything about these Peers.
%
%
Asymmetric communication is performed by having a Component send a message to another class's Component Id, or another object's Object Id. In this way client-server models are supported.

For these arrangements to work, different Peers must create Components with the same Object Ids. These Ids can either be agreed at design time, or more likely communicated via another channel, such as a Manager component.
For example, when Players create Avatars, they advertise an Object Id and a Prefab. The Avatar Manager creates instances at remote Peers with the correct Id.

The motivation for the two-part addressing is to isolate this synchronisation step from the user.
For example, Avatars define the Object Id, and all Components under the Avatar's scene branch use this Id (Figure~\ref{fig:addressing}). Only one Id needs to be synchronised for the entire branch to communicate.
A user could add a new Component to the Prefab to control eye-gaze. 
When the Avatar is instantiated at remote peers, communication between the new Components just works, because the Object Id has been set up by the Avatar Manager. 
The new Component class does not need to know about synchronising Object Ids, and the Avatar Manager does not need to know about the new class.

\paragraph{NetworkScene \& NetworkContext}

All networked Components are associated with a NetworkScene. This is a GameObject that interfaces between Components and the underlying Peer connections. It is analogous to the root of a scene graph.
When networked Components start, they find their parent NetworkScene and register themselves to it.
The NetworkScene parses message headers and is responsible for issuing callbacks on the correct Components.
The NetworkScene is also responsible for transmitting messages back over its connections. Components send messages through a NetworkContext convenience object that they receive when registering. This contains their identity (Object and Component Ids), and a reference to their NetworkScene.

For convenience, each NetworkScene has a unique Object Id. This allows Components implementing common services, such as Spawning and Avatar Management, to be placed below it.
Such Components can then address their counterparts at specific Peers by using that Peer's NetworkScene Id.
There is a one-to-one relationship between a Peer and a NetworkScene, though a \textit{process} may have multiple Peers \& NetworkScenes, each with their own set of connections.

A Component's NetworkScene is considered to be the first encountered descendent of their common ancestor.
The motivation for this, rather than using direct linage or a singleton pattern, is to provide flexibility in how scenes are built.
NetworkScene Prefabs with different Components, and so different capabilities, can be created and added to any scene, rather than making configurations project wide. 
Those Prefabs can be used in scenes with other Prefabs which may include networked Components, without cross-referencing or relying on variants.
It also allows use cases such as local loopback within a scene, as described in Section~\ref{sec:localloopback}.

\paragraph{The Medium is the Message}

Messages are only received by their intended Component(s). By virtue of having received a message, Components know how it was created, and so how it should be interpreted.
Messages are discrete binary blobs. Users can put anything inside these blobs, but are responsible for serialisation \& de-serialisation. 
Single-line methods to send and receive arbitrary objects as Json are included for simplicity. Json was chosen for its cross-platform support.

The motivation is for message exchange to be simple for new users, while remaining agnostic to serialisation so advanced users can choose appropriate methods.
For example, sending Avatar transforms might best be done through blitting, as these are fixed size, latency sensitive, but not too large.

\paragraph{Bootstrapping}

All Components, even those with a logical client-server model, run above the Ubiq messaging layer and assume that connections between Peers (NetworkScene instances) are already established.
NetworkScene instances create and manage the underlying connections, but user code must give the instructions to do so.
The Ubiq RoomClient Component (Section~\ref{sec:services}) is the only Ubiq Component that will create a connection on startup.
This design is a consequence of keeping the architecture separate from the messaging APIs.

\subsection{Services}
\label{sec:services}

Ubiq implements a number of common SVR features. These are designed with minimal interdependencies with the expectation that users may modify or replace them as needed.

\paragraph{XR Input}
A Player Controller is defined to allow navigation and interaction using desktop or common XR controls.
The code is designed with two separate, non-conflicting pathways, allowing the same Prefab to be used for desktop and XR with no changes. Ubiq includes a verb-based (use, grasp) system for 3D interaction. A component is provided to add 3D-ray support to world-space Unity Canvases, for easily constructing 2D UIs. This system uses the inbuilt Unity XR toolkit making it compatible with all Unity supported platforms.

\paragraph{Avatars}
An Avatar represents a player.
The Avatar Manager Component creates Prefabs to represent remote players based on their advertised avatar properties.
The default are stylised, floating avatars. In XR the head and hands are driven with a three-point tracking rig. On desktop they follow the direction of the camera.

\begin{figure}[h]
\includegraphics[width=8.5cm]{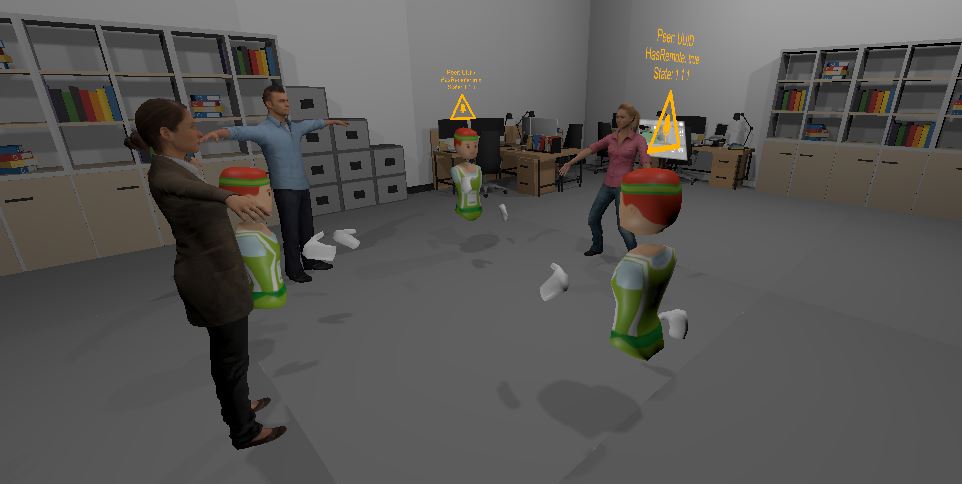}
\Description{A first person view of a virtual environment with a mixture of photo-realistic and stylised avatars standing in a circle, facing each other.}
\caption{Example of social gathering between multiple floating avatars and Microsoft RocketBox avatars, with overhead status indicators.}
\label{fig:rocketboxxamples}
\vspace{-0.3cm}
\end{figure}

There are no Prefab constraints other than that they have an Avatar Component at their base. It is straightforward to add additional Avatars with extended functionality.
For example, there are samples showing how to use the more realistic Microsoft RocketBox Avatars \cite{gonzalez-franco_rocketbox_2020}  
(Figure~\ref{fig:rocketboxxamples}).
The stylised Avatar includes a Component to change the texture. We have planned to support more social features by making avatars more expressive through control over facial expressions~\cite{tanenbaum_how_2020}. It is expected that developers will want to add additional Components to support extended customisability.
Avatar flexibility is important, not only to maximise potential modalities researchers may require, but the effects of mixed avatars is itself an active research topic~(e.g. \cite{latoschik_effect_2017}).

\paragraph{Voice Chat}
The Voice subsystem creates audio channels for real-time voice communication. Channels are established peer-to-peer according to the WebRtc specification. This allows cross-platform interaction with web-browsers, which only support WebRtc.
Ubiq uses a C\# implementation of WebRtc, rather than the Chromium library. This makes it easier to integrate with Unity's audio system, more transparent, and avoids the need to maintain platform-specific binaries.
Each {VoIP} channel is represented at either peer by a Component, created in code as-required.
In the samples, the {VoIP} Manager creates new Components as new Peers join a room.

APIs are provided to associate Avatars \& {VoIP} Components with individual Peers. These are used, for example, by status indicators that exist on the sample Avatars. The status indicators can show whether an avatar is speaking, as well as provide debugging information if a {VoIP} channel fails.

\paragraph{Object Spawning}

Ubiq provides APIs to instantiate Prefabs across all peers. This mechanism is used to instantiate the Avatars.

\paragraph{Event Logging}

The Event Logging subsystem is used for development and to instrument user experiments.
Users call a method to write events with arbitrary parameters, similar to other logging frameworks. However, these events can be transmitted across the network and collected at a single Peer (e.g. by an experimenter). Events can be collected post-hoc to support debugging experiments and applications.
User and application events are written as structured logs in Json. This allows easy ingestion by log aggregators, and tools such as Matlab, or Pandas in Python, for processing data programmatically.

\paragraph{Rendezvous and Rooms}

The Rooms system allows peers to join \textit{rooms} via secrets shared out-of-band. 
A room is a list of Peers forming a Peer network. All Peers in the network should exchange messages with each other.
The Rooms system is the only Ubiq component that has a logical client-server model. It assumes that the Peer network has one RoomServer and each Peer has one RoomClient. RoomServers and RoomClients have different Ids facilitating asymmetric exchange.

Though there is a logical client-server architecture, the Rooms system operates above the Ubiq messaging layer, so is still independent of the network architecture and requires bootstrapping.
The RoomClient will make the connection to a Peer hosting a RoomServer if a URI is specified.
Ubiq provides a C\# RoomServer to support local loopback, and a NodeJs RoomServer (Section~\ref{sec:server}) for production.

Users create rooms through the RoomClient. They receive a three-digit code for new rooms which can be shared out-of-band to allow other users to join.
The RoomClient features a number of events that are emitted as it and other Peers leave or join a room.
Components can register for these to implement behaviours such as automatically establishing voice chat to new peers, or creating avatars to represent them.
%

\paragraph{Scene Management}

Ubiq messaging is agnostic to the scene. An experimenter, for example, may want an overview scene with additional components for command and control.
The NetworkScene is designed to persist between scene changes, detaching the Scene from the Room.
Ubiq includes an optional component for changing scenes at runtime, allowing Peers in a Room to move through different scenes together.

\subsection{Server}

\label{sec:server}
Ubiq includes a server to facilitate rendezvous and room management. This is implemented on NodeJs. There is also an example C\# implementation within Unity itself (see Section \ref{sec:technical}).
The server accepts connections from Peers, over which it can exchange Ubiq messages.
Initially, the server will sandbox the connection. Messages addressed to the RoomServer are forwarded to the common RoomServer instance, while others are discarded.
Once the Peer negotiates membership of a room, the server will begin to forward messages between all the Peers in the room using the previously sandboxed connection(s).

\section{Examples}
\label{sec:examples}

\subsection{Social Example}

The \textit{Hello World} sample is intended to demonstrate Ubiq features, but itself functions as a complete SVR application, with UI for rendezvous, voice chat, avatars and interactive objects.

The example is made of four Prefabs: (1) the passive, static environment, (2) the NetworkScene Prefab with Manager Components for the various subsystems, (3) the Player Prefab which hosts the XR Input functionality, and (4) a Menu for driving the application. 

The Menu is a world-space Canvas with controls for browsing, creating and joining rooms. Additionally, a box is placed in the room that can spawn fireworks. The firework is intended as an introductory minimal working in-situ example of how to code shared objects.
The scene includes Components for all the services referred to in Section~\ref{sec:services}. The default behaviour of these services with regards to new peers means the scene acts as a fully functional meeting place, as well as a launchpad to other scenes.

\begin{figure}[h]
\includegraphics[width=8cm]{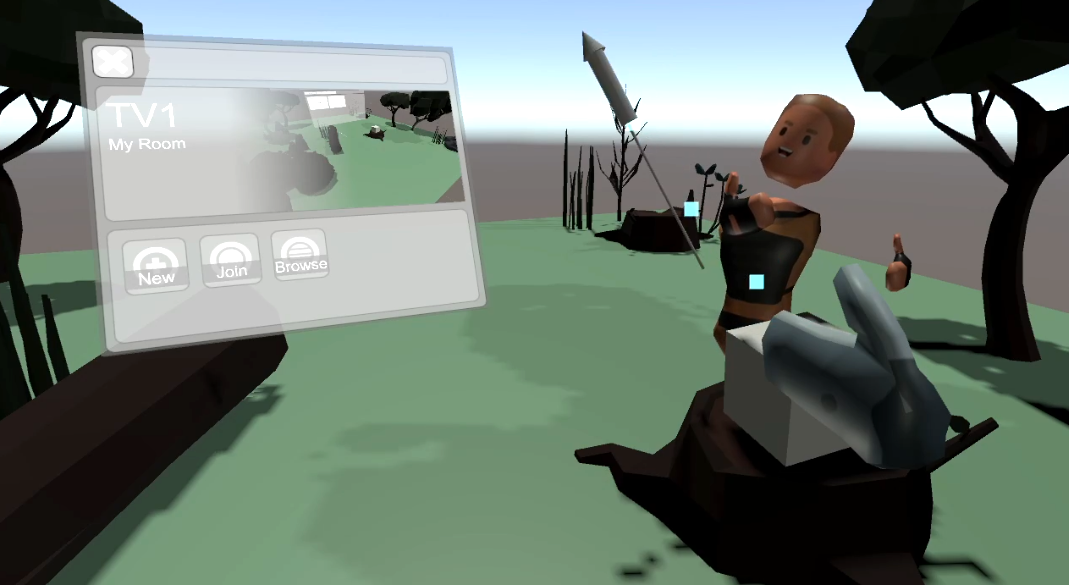}
\Description{First person view of the Ubiq Hello World scene, an outdoor space with logs and trees above a green ground plane. On the left is 2D menu embedded in 3D, with buttons to control room membership. To the right is another player's avatar, holding a firework and aiming it towards the sky.}
\caption{First-person view of the Hello World sample with a remote peer firing a Firework spawned from the interactive box. The Menu system includes controls for joining rooms with codes shared out-of-band.}
\label{fig:socialvrexample}
\vspace{-0.6cm}
\end{figure}

\subsection{Technical Examples}

\subsubsection{Local Loopback}
\label{sec:localloopback}

The Local Loopback sample (Figure~\ref{fig:technicalexamples}, left) shows two Peers running in a single Unity scene. This takes advantage of how Components find their NetworkScene (Section~\ref{sec:messaging}). As the search proceeds upwards, it is possible to create multiple `Peers' in one scene, simply by placing what would be the regular scene content under an empty GameObject. 
The NetworkScenes in each branch are independent; they could connect to a local server also in the scene, or to a production server. Connecting them all to the same room allows their Components to exchange messages as if they were on seperate machines, but in practice under one process.
This example shows how a developer can test distributed functionality, while having a single interface to trap events and debug code, and without having to set up a network or multiple machines.

\subsubsection{Boids}

\begin{figure}[h]
\includegraphics[width=8.5cm]{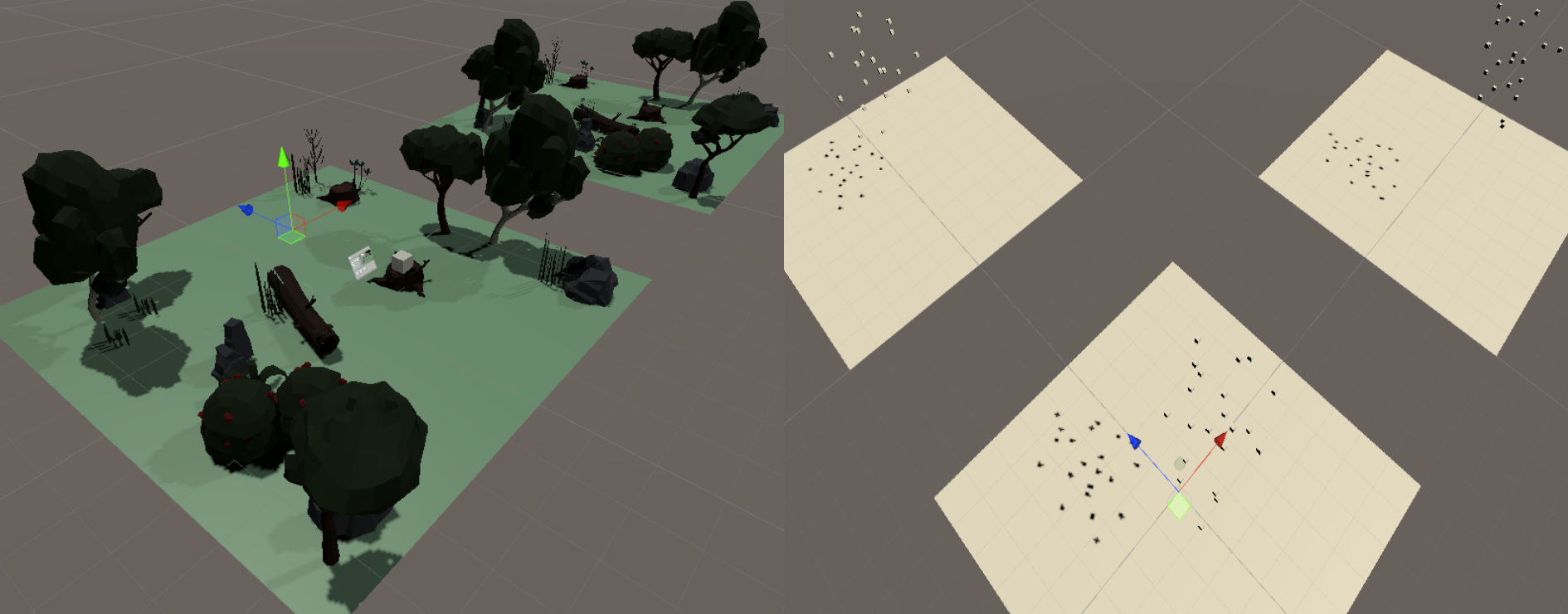}
\Description{On the left, a birds eye view of two instances of the Hello World environment, offset side by side in space. On the right, three instances of the 'boids' environment - a simple floor plane, with many floating cubes above it which move together in a flying flock. The flocks look the same in each instance suggesting they are synchronised.}
\caption{Local-loopback views of the Hello World SVR (left), and three Boids Peers sharing one flock with the same state (right). Each NetworkScene is offset in space.}
\label{fig:technicalexamples}
\vspace{-0.6cm}
\end{figure}

The Boids sample (Figure~\ref{fig:technicalexamples}, right) shows a flock of autonomous agents, where different agents are controlled by different Peers without a central authority.
Each Peer has a Boids Manager, which computes the state of its agents, based on the inertia of the whole flock.
The Manager exchanges its agents' states with instances on the other peers. Each peer has a copy of the entire flock, and peers calculate the same inertia independently to control their agents. 
Consequently, the flock moves collectively, even though control over individual agents is split between peers. This example provides a starting point to explore more complex simulations involving physics, prediction and compression.

\section{Evaluation}
\label{sec:evaluation}

\subsection{User-Level Functional Comparison}
\label{sec:userlevel}

Table~\ref{tab:taskcomparison} compares Ubiq's functionality with other SVRs using the task model from Liu \& Steed~\cite{liu_social_2021} along with their analysis of six well-known social VR systems: VRChat \cite{vrchat_2020}, RecRoom~\cite{rec_room_rec_2020}, AltSpaceVR~\cite{microsoft_altspacevr_2020}, BigScreen~\cite{bigscreen_inc_bigscreen_2020} , Spatial~\cite{spatial_systems_inc_spatial_2020}, and Mozilla Hubs~\cite{mozilla_corporation_hubs_2020}.
Their model breaks down tasks, such as finding other users, into \emph{interaction cycles} that classify how the system is used to achieve a goal - or whether it is even possible.  The interaction cycles used in Table~\ref{tab:taskcomparison} are explained below. Some tasks require two cycles, e.g. "2D + Switch" indicates both 2D and out-of-VR interaction are necessary. Others can be achieved in different ways, e.g. "3D/2D" suggests the user can choose between 2D or 3D interaction for that task.

\begin{itemize}
    \item 2D - interacting with 2D interfaces, e.g. menus
    \item 3D - interacting with 3D objects
    \item Goal - searching for a target in the environment
    \item Exploratory - exploring the environment for understanding
    \item Collaboration - interacting with other users
    \item System - system notifications are displayed
    \item Switch - action can only be completed outside VR
\end{itemize}

Ubiq supports most of the common SVR subtasks. 
The biggest differences come from tasks that require the system to have the concept of Peer groups, e.g. Group Moving. As Ubiq does not have user accounts there is no system for building friends lists. Users must share join codes out-of-band to meet in a room. 
%
Accounts could be added to Ubiq without significant changes. It would require adding a UI to login to an external service, and updating the local Peer with persistent data from it. Both tasks would be quite straightforward.

Ubiq includes a 2D menu system for creating public and private rooms. Users communicate through voice chat, and by gesturing with the head and hands, but there is no system to control facial expressions.
Because it is open source, Ubiq supports extending fidelity at a functional level, which is not possible in other systems that may allow different 3D models, but only on a fixed rig.

Ubiq lacks other features that some of these platforms support. For example, it has few built-in tools for scene modification, inventories of objects, libraries of games, etc. While it is primarily targeted at research and teaching, there are no particular restrictions in building some of these functionalities on top of Ubiq. As we will see in Section \ref{sec:teaching}, as the system is based around Unity, it is relatively easy to build game-like experiences of high quality.\\

\begin{table*}[]
\centering
\scriptsize
\vspace{-0.2cm}
\caption{Task-based functional comparison}
\vspace{-0.4cm}
\begin{tabular}{lllllllll}\toprule
Task & Subtask & VRChat & RecRoom & AltSpaceVR & BigScreen & Spatial & Mozilla Hubs & Ubiq \\
\midrule
Identification & Identify Others & Explore & Explore & Goal + 3D & Explore & Explore & Explore & 2D \\
& Identify Speaker & Explore+2D & Explore & Explore + 2D & Explore + 2D & Explore + 2D & Explore & 2D \\
& Identify Interactor & Explore & 3D & Explore & Explore & Explore & Explore & Explore \\
Communication & Express Emotion & 2D & 2D & 2D & - & - & 2D & - \\
& Gesture & Collaboration/2D & Collaboration & Collaboration & Collaboration & Collaboration & Collaboration & Collaboration \\
& Mark Friends & 2D + Switch & 2D + Switch & 2D + Switch & - & - & - & - \\
& Text & 2D + System & 2D + System & 2D & - & - & System & - \\
Navigation & Group Gather & Goal & Goal & Goal & Goal & Goal & Goal & Goal \\
& Group Moving & Goal/2D & 2D & 2D/Goal & 2D & 2D & - & - \\
& Room Transport & 2D & 2D/Switch & 2D & 2D/Switch & 2D/Switch & Switch & Switch \\
Manipulation & Create Objects & - & 2D + 3D & - & 2D & 2D & 2D & 2D/3D \\
& Move Objects & 3D & 3D & 3D & 3D & 3D & 3D & 3D \\
& Pass Objects & Collaboration & Collaboration & Collaboration & Collaboration & Collaboration & Collaboration & Collaboration \\
Coordination & Create Room & 2D & 2D & 2D & 2D & 2D & - & 2D \\
& Invite Others & 2D + System & 2D/Switch & 2D + System & Switch & 2D + Switch & Switch & Switch \\
& Public Room Meeting & 2D + Goal & 2D + Goal & 2D + Goal & 2D + Goal & - & - & 2D + Goal \\
& External Source Sharing & - & - & - & Switch & Switch & - & - \\
& System Notification & System & System & System & - & - & - & System \\
\bottomrule
\end{tabular}
\label{tab:taskcomparison}
\end{table*}

\subsection{Technical Comparison}
\label{sec:technical}

\begin{table*}[]
\centering
\scriptsize
\caption{Technical Comparison to other Unity frameworks. (*)All solutions would support a third-party voice solution such as Dissonance~\cite{dissonance}, but the solution would not be open source. (**)Users must create their own server code using the framework APIs.}
\vspace{-0.5cm}
\begin{tabular}{llllll}\toprule
& Photon & DarkRift & UNet/Mirror & MLAPI & Ubiq \\
\midrule
Network Architecture & Client-Server & Client-Server & Client-Server & Client-Server & Flexible \\
API Level & Low-mid level & Low-mid level & Low-level & Low-level & Low-mid level \\
Voice & Photon Voice & No* & No* & No* & Built-In \\
Transport & Own & Options & Various integrations & Unity.Transport (options) & Various, inc. WebSockets \\
Platforms & Unity & Unity & Unity & Unity & Unity/NodeJs/Browser \\
SocialVR & Additional & No & No & No & Out-Of-Box \\
Client Licence & Closed & Commercial Source &  MIT & Unity & Apache \\
Service Cost & Rental & Self-Hosted & Self-Hosted** & Self-Hosted** & Free/Self-Hosted \\
\bottomrule
\end{tabular}
\label{tab:technicalcomparison}
\end{table*}


Table~\ref{tab:technicalcomparison} compares Ubiq with some of the more mature networking frameworks that are available for Unity. Photon~\cite{photon} and DarkRift~\cite{darkrift} are commercial solutions but have free services for smaller projects. UNet is a Unity built-in networking package that was deprecated, but revived as Mirror~\cite{mirror}. MLAPI~\cite{mlapi} is the recent Unity Multiplayer API. At the time of writing this was still experimental. These other frameworks are low-to-mid level. This may be because creating high level networking APIs requires too many strict assumptions which are incompatible with a flexible tool such as Unity. While Ubiq is designed primarily for Unity, it has support for cross-platform play with web browsers. Being fully open source, Peers could be written for other platforms too.

Ubiq is distinct in that its client-server architecture is not inherent. If an ownership/host model is used it must be implemented in user code. All other systems, even the unhosted ones, assume a client-server architecture with processes themselves identifying as clients or hosts/servers. In Ubiq, different Components within the same application can use different models.


Voice support is also a core feature. The only other framework to provide this is Photon. Ubiq uses the open WebRtc standard for negotiating peer-to-peer connections, providing full transparency of the audio dataflow.

Ubiq includes a NodeJs server. The server is not necessary for messaging, but supports rendezvous and matchmaking for the Rooms system and social samples. Most other frameworks require a server to be written by the user. This will be less effort than it sounds however, as the APIs have the client-server model inbuilt, so server functionality will consist mainly of gameplay logic. Photon provides the option to rent a managed server. We provide free public access to our development servers, and the source allowing users to self-host if they prefer.

\subsection{Performance}

The bandwidth of an SVR is dominated by developer decisions about update frequency. This is flexible in Ubiq, but it interesting to understand the latency and overhead with Ubiq's current conventions. We captured a typical SVR session with 8 users lasting 22 minutes to characterise the network behaviour. We recorded the latencies between all peers, and the throughput at one peer. This also serves as a demonstration of Ubiq's built-in instrumentation. 

\begin{figure}[h]
\includegraphics[width=7.0cm]{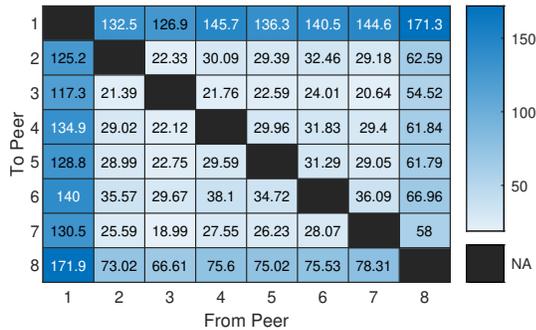}
\Description{A grid of cells showing the latencies from all peers to all other peers. The peer Ids are along the x and y axes. The cells contain the latencies from the corresponding peer on the y axis to the corresponding peer on the x axis. The diagonal cells where the to and from peers are the same are blanked out. The lower-left and upper-right regions of the grid are generally symmetrical along the diagonal, suggesting the ping times between two peers are similar regardless of which issues the ping. The times range between 30-130 ms suggesting the ping time is dependent on the peer.}
\caption{Average latencies (in ms) between all peers in the session.}
\label{fig:latencies}
\vspace{-0.3cm}
\end{figure}


Peer-to-peer latencies are shown in Figure~\ref{fig:latencies}. Latencies were sampled continuously at 1 Hz, by each peer to all others, measured as half the round trip time.
The eight peers were split across three regions on a continent (A,B,C). The distances between the nearest cities were 
260~km (A-B),
1250~km (B-C),
1000~km (C-A).
Two peers were on the same local network.

\begin{figure}[h]
\includegraphics[width=7.5cm]{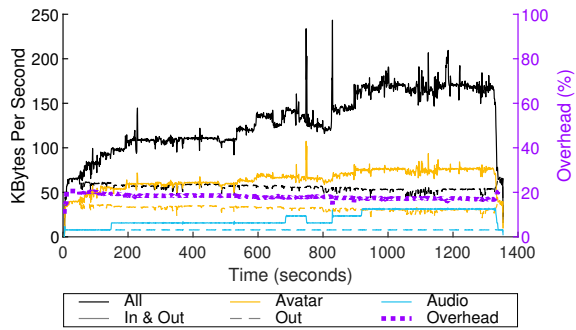}
\Description{A line graph showing bandwidth in kilobytes per second against time, between 0 and 1400 seconds, and a second line graph showing overhead in percent, against the same time.
There are six bandwidth series, which break down bandwidth into different groups, including the total bandwidth, and only bandwidth for audio and avatar data. There is one series for overhead. The overhead remains constant at 20\% for the entire time. The bandwidth fluctuates, peaking steadily around 1000 seconds when the largest numbers of peers are in the room. The bandwidth series are always proportional to each other.}
\caption{Throughput \& Overhead of the desktop client. Measurements include the bandwidth consumed by the latency measurements.}
\label{fig:throughput}
\vspace{-0.3cm}
\end{figure}

Figure~\ref{fig:throughput} shows the throughput and messaging overhead on one client. The SVR is symmetrical, so the throughput should be proportionally representative of all peers. For performance reasons though, throughput was captured on a desktop peer, which had a higher update rate and so higher transmission rate than the others. Figure~\ref{fig:throughput} shows the overall bandwidth, as well as a breakdown of the bandwidth dedicated to avatar data and voice data.

The step changes in bandwidth as new peers join can be seen, especially in the audio, which uses the fixed-rate G722 codec. At the busiest point bandwidth is stable between 150-200 Kbyte/s. The largest proportion of the bandwidth is used for the avatar data.

Ubiq messages are prefixed by a length and address, a total of 14 bytes. 
Audio channels establish their own per-dyad connections, so do not have this overhead.
%
Figure~\ref{fig:throughput} shows the overhead of the prefix as a proportion of the Ubiq message bandwidth.
Though this will vary with the type of data being sent (the message length), in the capture session it was very stable at approximately 20\%.

To measure capacity, we created a bot to emulate a user, and connected increasing numbers of bots to a room while monitoring performance.
The QoE at the client was approximated by FPS, and the QoS of the server by the peer-to-peer latencies through it.
We found the server (dual-2.2 GHz/4GB RAM VM) could handle 50 users in a room before significant increases in latency were observed. A client running on a 2.5GHz/Embedded Graphics desktop could support 30 peers before the FPS dropped below the native rate (60).

\subsection{Teaching Use}
\label{sec:teaching}

Ubiq was used as a basis for the coursework of our Virtual Environments module in the teaching year 2020/2021. Students worked in groups of 3-5 to develop an SVR environment that required several users to collaborate or compete. Teaching was remote, so a key motivation was that students could meet in the SVR as a group. 
Ubiq supplied common features, and examples on implementing shared interactive objects, allowing students to focus on building interesting interactions. The ability of Ubiq to work with and without XR was important as it was not possible to supply all students with VR headsets due to their geographic spread.

The students worked on this project part-time (20\%) for two months under regular supervision, and developed a range of applications. 
It was left up to them whether the SVR would be a game or serious work environment, but naturally, most applications were the former.
Of sixteen projects developed, two are described here to provide examples of what the students were able to realise.

\begin{figure}[h]
\includegraphics[width=8.5cm]{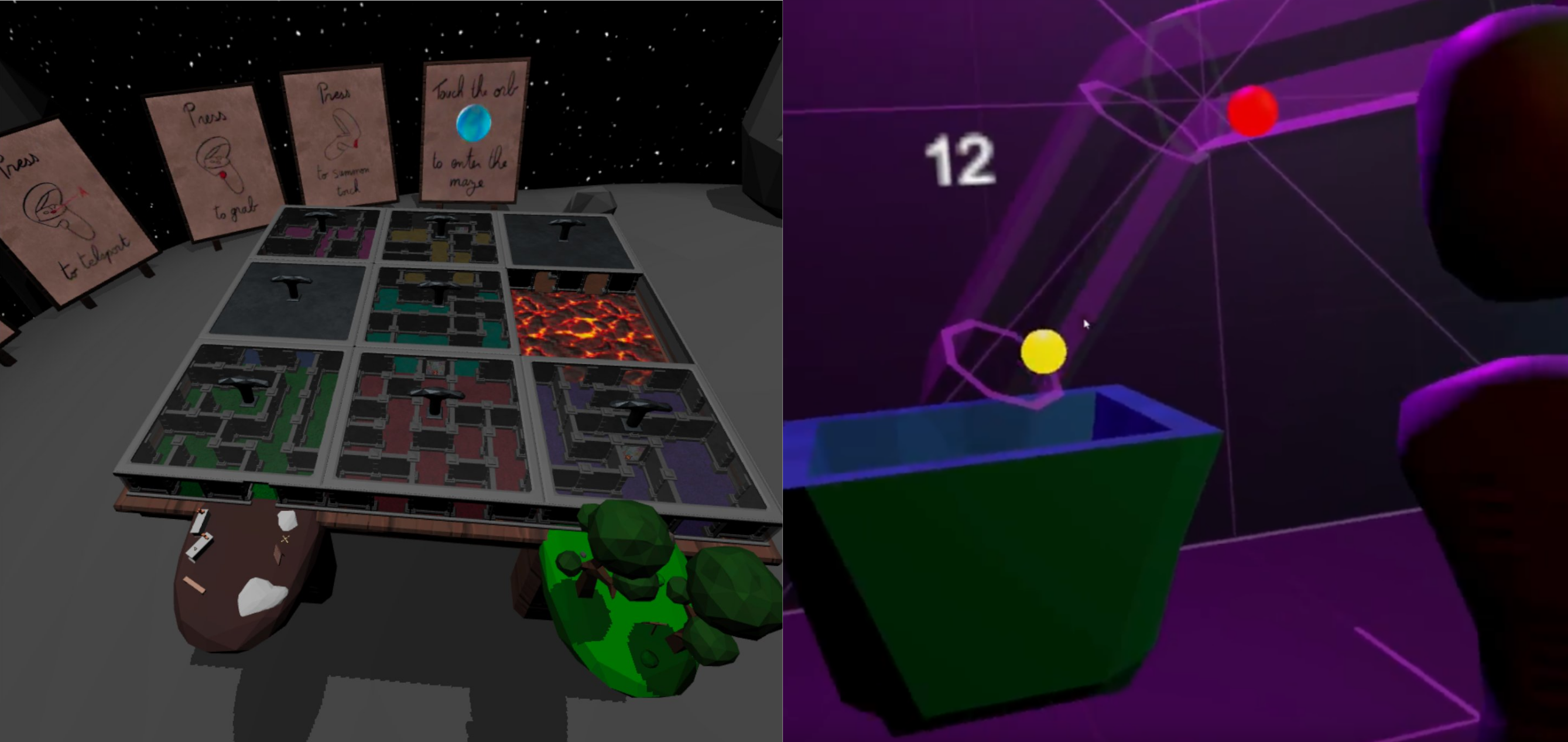}
\Description{On the left, a first person view of the table maze described for Two-Scale Maze, showing capped and uncapped tiles, and the starting and ending points for the small-scale players. These are small platforms attached to the maze, both showing outdoor scenes, but one barren and one lush. On the right, a first-person view inside Transballer, showing bright coloured balls falling down a player-positioned chute into a hopper.}
\caption{Example student projects: view of the normal-scale player in Two-Scale Maze (left) and of balls being fed into the hopper in front of another player in Transballer (right).}
\label{fig:studentexamples}
\vspace{-0.3cm}
\end{figure}

\paragraph{Two-Scale Maze} Two-Scale Maze (\autoref{fig:studentexamples}, left) is a combination of a sliding puzzle and a labyrinth. Players must arrange tiles on a table to assemble a maze that can be traversed. However, some tiles are covered, so the only way to see how they should fit together is for at least one of the players to shrink to a fraction of their size and traverse the incomplete maze while communicating to the full-size players what needs to be changed.

\paragraph{Transballer} Transballer (\autoref{fig:studentexamples}, right) is a collaborative take on the \textit{Fantastic Contraption}-style game.
Players begin in an empty room with a spawn point and a goal. The challenge is to use assorted items in their inventory to create a scaffolding to carry balls across the room. The challenge can be increased with the addition of obstacles and multiple spawn points.

These applications demonstrate decidedly non-trivial functionality, which would be difficult to implement on some platforms.
Two-Scale Maze has interaction take place at different scales. Not only did the messaging have to support scale transitions, but many in-built systems such as physics and lighting assume a uniform scale. To overcome this challenge the team used two different scale scenes and leveraged Ubiq's agnostic messaging to create the impression that users were in the same scene.
Transballer had to support large numbers of shared physical objects controlled by multiple users. 
Keeping a physics simulation consistent across multiple clients is a complex problem which the team addressed by creating a comprehensive per-object peer-to-peer ownership model.

These demonstrations, and others, will be made available as a showcase of work in Ubiq. Some of the students have chosen to make their projects open source.


\section{Future Work}

We have planned additional features and experiments to perform with Ubiq.
To improve accessibility for researchers, we will add additional tutorials on common use cases, and provide a number of real distributed experiments, including our In-The-Wild presence experiment \cite{Steed2016} and collaborative embodiment experiment~\cite{Pan2017}.
%
%
To expand the feature-set and potential use cases, we will enhance the rooms system with new models for scalability, and improve the in-built avatar customisation. We will add a demonstration of persistent accounts to support groups.
We are also developing cross-platform support further, including a web client example. In the first instance this will be a visualiser that can perform command and control, though as Ubiq's messaging is platform agnostic there is nothing preventing it becoming a fully symmetrical peer. Additionally, examples of procedural clients that can connect to provide data from external sources will be added. Our aim is to have Ubiq synchronise users across multiple heterogeneous AR \& VR devices.


\section{Conclusions}
\label{sec:conclusions}

We present Ubiq, a system for building cross-platform SVR in Unity.
Ubiq was created to fulfil goals that are unlikely to be addressed by commercial SVRs, owing to conflicts with their business model.
These include being open source to support new features, the ability to self-host for data protection and cost concerns, and being agnostic to the network architecture to support network research. This is in addition to services such as logging and remote code execution which are often not exposed in commercial systems.

Ubiq is foremost a framework. Its expected use case is for small teams to use it to build their own applications.
As a starting point, a fully functional SVR sample is provided out of the box.

Functionality-wise Ubiq lacks features pertaining to persistent users, such as the ability to build friend-groups. However it has a number of services that other systems are missing or do not expose, such as logging and instrumentation.
Ubiq is most distinct in that it decouples messaging from network architecture. Users control the authority and routing models themselves through Ubiq's addressing scheme. Importantly, they can use different models on a per-Component basis.
Ubiq does include a server. This is not necessary for Ubiq to function, but is a practical requirement for rendezvous over the public internet.

Ubiq meets the framerate goals of platforms such as the Oculus Quest and shows relatively low bandwidth consumption in small group meetings.
Ubiq has been tested in a classroom setting and used successfully by students without prior networking experience. 

Many toolkits have been presented over the years.
However VR applications are moving away from integrated platforms towards commercial game engines.
Ubiq has been built for this new practice, adapting lessons from previous designs to modern tools, and maximising the use of standards and integrations.
The design scope is kept deliberately large to capture niche use cases.
Ubiq has a number of internal and external users with substantial projects to maintain momentum, and we make our documentation available to other teaching teams.
Ubiq is available at \url{https://github.com/Ubiq/Ubiq} under the permissive, commercial friendly Apache license.

\bibliographystyle{ACM-Reference-Format}
\bibliography{sample-base}

\end{document}